# DB-GNN: Dual-Branch Graph Neural Network with Multi-Level Contrastive Learning for Jointly Identifying Within- and Cross-Frequency Coupled Brain Networks

Xiang Wang
*ACS Lab*
*Huawei Technologies*
Shanghai, China
wangxiang224@huawei.com

Hui Xu
*Joint Laboratory of Bioimaging Technology and Applications*
*SAS-SIM-IT & MEDI*
Shanghai, China
hxu@cshmedi.com

Jing Cai
*School of Health Technology and Informatics*
*the Hong Kong Polytechnic University*
Hong Kong, China
jing.cai@polyu.edu.hk

Ta Zhou
*School of Computing*
*Jiangsu University of Science and Technology*
Zhenjiang, China
jkdzhout@just.edu.cn

Xibei Yang
*School of Computing*
*Jiangsu University of Science and Technology*
Zhenjiang, China
jsjxy_yxb@just.edu.cn

Wei Xue
*Academy for Engineering & Technology*
*Fudan University*
Shanghai, China
weixuejkd@163.com

*Abstract*—Within-frequency coupling (WFC) and cross-frequency coupling (CFC) in brain networks reflect neural synchronization within the same frequency band and cross-band oscillatory interactions, respectively. Their synergy provides a comprehensive understanding of neural mechanisms underlying cognitive states such as emotion. However, existing multi-channel EEG studies often analyze WFC or CFC separately, failing to fully leverage their complementary properties. This study proposes a dual-branch graph neural network (DB-GNN) to jointly identify within- and cross-frequency coupled brain networks. Firstly, DB-GNN leverages its unique dual-branch learning architecture to efficiently mine global collaborative information and local cross-frequency and within-frequency coupling information. Secondly, to more fully perceive the global information of cross-frequency and within-frequency coupling, the global perception branch of DB-GNN adopts a Transformer architecture. To prevent overfitting of the Transformer architecture, this study integrates prior within- and cross-frequency coupling information into the Transformer inference process, thereby enhancing the generalization capability of DB-GNN. Finally, a multi-scale graph contrastive learning regularization term is introduced to constrain the global and local perception branches of DB-GNN at both graph-level and node-level, enhancing its joint perception ability and further improving its generalization performance. Experimental validation on the emotion recognition dataset shows that DB-GNN achieves a testing accuracy of 97.88% and an F1-score of 97.87%, reaching the state-of-the-art performance.

## I. INTRODUCTION

Electroencephalography (EEG) has proven to be an invaluable tool for understanding the intricacies of brain function due to its ability to capture both high-temporal-resolution neuronal oscillations in local brain areas and broader patterns of synchronization between different regions [1]. These properties make EEG highly suitable for investigating the neural basis of complex phenomena such as emotions, which are inherently dynamic and involve complex interactions across brain networks [2].

The potential of multi-channel EEG in emotion recognition lies largely in its capacity to capture both within-frequency coupling (WFC) and cross-frequency coupling (CFC), which represent two distinct yet interconnected aspects of neural communication [3]. The WFC reflects neural activity within the same frequency band, while the CFC captures the interplay between different frequency bands, both of which provide insights into how neural processes at different temporal scales interact to support cognitive and affective states [4]. WFC has been associated with the processing of external sensory events [5], the internal attentional selection [6], and the emotional states [7]. The CFC has been shown to contribute to the integration of information that leads to memory consolidation [8], and to emotion recognition due to its enriched information [9].

As shown in Fig. 1, previous studies utilizing EEG for emotion recognition have primarily focused on either WFC or CFC in isolation, without investigating the interactions between these two forms of coupling [10]. Cui et al [7] demonstrated that differences exist in WFC when subjects are in different emotional states, while Zhang et al [9] showed that Granger causality features in CFC can further improve the recognition accuracy of emotion states. However, the simultaneous exploration of both WFC and CFC (Fig. 1), and how their interplay contributes to a more comprehensive understanding of emotional processing, has been largely overlooked [11]. This gap in the literature presents an opportunity to develop a more integrated approach that can capture the full spectrum of neural dynamics underlying emotion recognition [12].

On the other hand，when analyzing the EEG signals identifying the topological patterns of brain networks from multi-channel EEG signals, traditional methods rely on graph theoretical metrics for classification, such as node degree, clustering coefficient, average path length, hubs, centrality, modularity, robustness, and assortativity [12], which can serve as indicators of cognitive and behavioral performance. Recent development in deep learning architecture has inspired the network neuroscience community to utilize artificial neural networks to extract and classify multi-channel EEG signals in an end-to-end manner. Compared with traditional methods based on graph theoretical metrics, graph neural networks (GNNs) replace the manual feature engineering with representational learning and have proved their advantages in capturing both spatial and temporal features from neurophysiological data [13]. However, previous work [13] building GNNs by neighborhood aggregation may fail to capture complex global features due to their relatively small receptive field, and those combining the Transformer structure with GNNs [14], even equipped with the ability of global reception, may suffer from overfitting due to the overparameterization and fail in its generalization to different brain networks . Consequently, this study proposes a dual-branch graph neural network that incorporates the local information from each WFC/CFC brain network and the global information from all WFC/CFC brain networks.

This study proposed a dual-branch graph neural network (DB-GNN) for EEG emotion recognition (Fig. 2). First, EEG signals are decomposed into five different frequency bands. To measure different emotion brain states, phase locking value (PLV) [7] is calculated within-frequency band to construct the WFC brain networks and modulation index (MI) [16] is computed between pairs of frequency bands to generate the CFC brain networks. Subsequently, under the constraint of the prior coupling information, a Transformer-based GNN model is proposed to perceive the global topology of brain networks. Moreover, we employ multi-level graph contrastive learning to regularize the learning process of the model, enhancing its generalizability. The contributions of our work are summarized as follows.

1. **Dual-branch learning architecture**：DB-GNN employs a dual-branch learning architecture, utilizing Graph Attention Networks (GAT) [17] to separately extract local features from within-frequency and cross-frequency coupled brain networks, and leveraging a prior information-based graph transformer module (PiGTM) to capture collaborative global features across all brain networks. This approach fully leverages both WFC and CFC information of emotional brain states, thereby laying a foundation for accurate emotion classification.

2. **Prior information-based graph transformer module (PiGTM)**：To enhance the global feature perception capability of DB-GNN, this study proposes employing a Transformer architecture to enable collaborative global information perception across all WFC/CFC brain networks. To mitigate the risk of overfitting, prior coupling information of brain networks was integrated into the self-attention learning mechanism, thereby constraining the feature perception process and enhancing the model's generalization capability.

3. **Multi-level graph contrastive learning**：The DB-GNN incorporated both graph-level and node-level graph contrastive learning to constrain the dual-branch learning architecture, thereby enhancing the ability of DB-GNN to collaboratively perceive global and local features of brain networks while mitigating the overfitting risk associated with considering only local or global features in isolation.

The remaining sections of this paper are organized as follows. Section 2 summarizes related works. Furthermore, section 3 details the construction of DB-GNN. Subsequently, the comparative experiment and analysis is arranged in section 4. Finally, section 5 concludes this study.

## II. Related Works

Research in analyzing brain networks for brain state classification can be categorized into two different directions: (1) Graph theoretical metrics, such as node degree, clustering coefficient, average path length, hubs, centrality, modularity, robustness, and assortativity [12], are extracted from both the WFC and CFC brain networks [18] and form the feature basis for recognition tasks. (2) Artificial neural networks from the deep learning community, such as the convolutional neural network (CNN) and the GNN, are deployed to resolve the classification of brain states in an end-to-end manner [14]. These studies highlighted the fundamental role of WFC and CFC in brain state classification. However, a systematic way of combining the information from both WFC and CFC for analyzing brain networks is still lacking [11], and the potential of this direction in brain state classification tasks such as emotion recognition is worth further exploration.

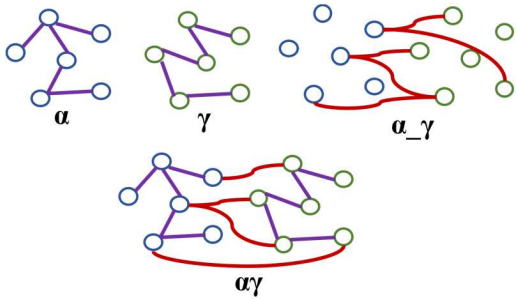

Fig. 1. Within-frequency and cross-frequency coupling brain networks. $\alpha$ and $\gamma$ denote the within-frequency coupling brain networks constructed from multi-channel EEG signals in the $\alpha$ and $\gamma$ frequency bands, respectively. $\alpha\_\gamma$ represents the cross-frequency coupling brain network between the $\alpha$ and $\gamma$ frequency bands. Previous studies analyze $\alpha$, $\gamma$, or $\alpha\_\gamma$ brain networks separately. This study seeks to jointly extract the global collaborative features of all cross-frequency and same-frequency coupling brain networks (i.e., $\alpha\gamma$).

Designed for operating on graph-structured inputs, GNNs have emerged as a successful tool in a variety of fields, such as natural language processing [19], bioinformatics [20], and network neuroscience [21]. Recently, GNNs have also been introduced for EEG-based emotion recognition [14], [21]. For example, several works use GNNs to capture both local and global relations among different EEG channels for better emotion recognition [22]. Moreover, Guo et al [14] proposed a dynamic graph neural network model and utilized the self-attention mechanism to enhance the representational power of the output. Fan et al [13] proposed a RGNet based on a novel region-wise encoder and obtained an average recognition accuracy of 98.64% and 99.33% for Deap and Dreamer datasets, respectively. These results show the promising potential of obtaining graph structural information in improving the performance of emotion recognition models.

Graph contrastive learning methods [23] have been successfully applied for learning robust and discriminative representations from graph-structured data by leveraging self-supervised learning. These methods aim to maximize the agreement between positive pairs of graph representations while minimizing the similarity to negative samples, typically using contrastive loss functions such as InfoNCE [23]. Graph contrastive learning techniques have shown great promise in enhancing the robustness of emotion recognition models based on multi-channel EEG signals. For instance, Gilakjani et al [24] employed graph contrastive learning to improve the quality of the learned representations from EEG signals and mitigate the adverse impact of inter-subject and intra-subject variability in signals corresponding to the same stimuli or emotions. Zhang et al [25] utilized graph contrastive learning to facilitate their synchronous training of graph convolutional networks on multisubject datasets. Given the power of the graph contrastive learning in addressing overfitting and data scarcity challenges, we decided to take advantage of these methods to constrain the training process of our proposed DB-GNN model.

III. METHODS

A. Construction of Brain Networks

Suppose that the raw multi-channel EEG signals are $\{\mathbf{X}_i \in \mathbb{R}^{1 \times N}\}_{i=1,2,\ldots,C}$, with $\mathbf{X}_i$ representing the $i$-th channel with $N$ samples and $C$ representing the number of channels. First, we used Butterworth filter [26] to extract 5 different frequency bands from the multi-channel EEG signals: $\delta$ (1-3Hz), $\theta$ (3-8Hz), $\alpha$ (8-12Hz), $\beta$ (12-30Hz) and $\gamma$ (30-48Hz):

$$\mathbf{X}_i^\kappa = \psi_\kappa(\mathbf{X}_i), \kappa \in \{\delta, \theta, \alpha, \beta, \gamma\} \quad (1)$$

To quantify the WFC of multi-channel EEG signals, this study used the phase locking value (PLV) [7] to measure the strength of phase-phase coupling between different EEG channels in the same frequency band. Meanwhile, modulation index (MI) [16] is utilized to measure the strength of phase-amplitude coupling across frequency bands. The PLV is calculated as

$$PLV_{ij}^\kappa = \frac{1}{N}\left|\sum_{n=1}^{N} exp\left(i\hbar_i(\varepsilon,n) - i\hbar_j(\varepsilon,n)\right)\right|$$
$$\kappa \in \{\delta, \theta, \alpha, \beta, \gamma\} \quad (2)$$

where $PLV_{ij}^\kappa$ represents the value of PLV in the $\kappa$ frequency band between channel $i$ and channel $j$, with $i,j = 1,2,\ldots,C$. $\{\hbar_i(\varepsilon,n)\}_{n=1,2,\ldots,N}$ represents the Hilbert Transform of the EEG signals from channel $i$ with $N$ sampling points and length $\varepsilon$. The MI is computed as

$$MI_{ij}^{\kappa_1\kappa_2} = \frac{KL(U,X)}{logN}, KL(U,X) \leftarrow \left(logN - H(p)\right) \quad (3)$$

where

$$H(p) = -\sum_{j=1}^{N} p(j)\log p(j), p(j) \leftarrow \frac{\bar{a}}{\sum_{k=1}^{N} a_k} \quad (4)$$

$\bar{a}$ is the average amplitude of a single time bin. $N$ is the total number of bins.

Based on the above equations, we constructed the adjacency matrix that quantifies the CFC and WFC of the brain network as

$$\mathcal{A} = \begin{bmatrix} \mathbf{A}^\delta & \mathbf{A}^{\delta\theta} & \mathbf{A}^{\delta\alpha} & \mathbf{A}^{\delta\beta} & \mathbf{A}^{\delta\gamma} \\ \mathbf{A}^{\delta\theta} & \mathbf{A}^\theta & \mathbf{A}^{\theta\alpha} & \mathbf{A}^{\theta\beta} & \mathbf{A}^{\theta\gamma} \\ \mathbf{A}^{\delta\alpha} & \mathbf{A}^{\theta\alpha} & \mathbf{A}^\alpha & \mathbf{A}^{\alpha\beta} & \mathbf{A}^{\alpha\gamma} \\ \mathbf{A}^{\delta\beta} & \mathbf{A}^{\theta\beta} & \mathbf{A}^{\alpha\beta} & \mathbf{A}^\beta & \mathbf{A}^{\beta\gamma} \\ \mathbf{A}^{\delta\gamma} & \mathbf{A}^{\theta\gamma} & \mathbf{A}^{\alpha\gamma} & \mathbf{A}^{\beta\gamma} & \mathbf{A}^\gamma \end{bmatrix} \quad (5)$$

with $\mathbf{A}^\kappa \leftarrow (a_{ij}^\kappa) \in \mathbb{R}^{C \times C}$ and $\mathbf{A}^{\kappa_1\kappa_2} \leftarrow (a_{ij}^{\kappa_1\kappa_2}) \in \mathbb{R}^{C \times C}$ constructed as following:

$$a_{ij}^\kappa = \begin{cases} 1, & PLV_{ij}^\kappa \geq \tau_1 \\ 0, & PLV_{ij}^\kappa < \tau_1 \end{cases}, \kappa \in \{\delta, \theta, \alpha, \beta, \gamma\}, i,j = 1,2,\ldots,C \quad (6)$$

$$a_{ij}^\kappa = \begin{cases} 1, & MI_{ij}^{\kappa_1\kappa_2} \geq \tau_2 \\ 0, & MI_{ij}^{\kappa_1\kappa_2} < \tau_2 \end{cases}, \kappa_1 \neq \kappa_2 \in \{\delta, \theta, \alpha, \beta, \gamma\}, i,j = 1,2,\ldots,C \quad (7)$$

where $\tau_1$ and $\tau_2$ are manually set thresholds, which are chosen to keep 20% of the brain network density.

B. PiGTM: Prior information-based graph transformer module

Let $\mathcal{G} = (\mathcal{V}, \mathbf{A})$ represents the brain network, where $\mathcal{V} = \{v_1, v_2, \ldots, v_n\}$, $n = |\mathcal{V}|$ is the number of nodes. $\mathbf{A}$ represents the adjacency matrix of the brain network. Let the feature vector of node $v_i$ be $\mathbf{h}_i \in \mathbb{R}^{1 \times d}$. The feature matrix $\mathbf{H} \leftarrow (\mathbf{h}_1; \mathbf{h}_2; \ldots; \mathbf{h}_n) \in \mathbb{R}^{n \times d}$ is mapped to the corresponding representation **Q**, **K**, **V** (Fig. 2) by three matrices $\mathbf{W}_Q \in \mathbb{R}^{d \times d_K}$, $\mathbf{W}_K \in \mathbb{R}^{d \times d_K}$ and $\mathbf{W}_V \in \mathbb{R}^{d \times d'}$, where $d_K$ is Q, K's dimension:

$$\mathbf{V} = \mathbf{H}\mathbf{W}_V \in \mathbb{R}^{n \times d'} \tag{8}$$

$$\mathbf{Q} = \mathbf{H}\mathbf{W}_Q \in \mathbb{R}^{n \times d_K} \tag{9}$$

$$\mathbf{K} = \mathbf{H}\mathbf{W}_K \in \mathbb{R}^{n \times d_K} \tag{10}$$

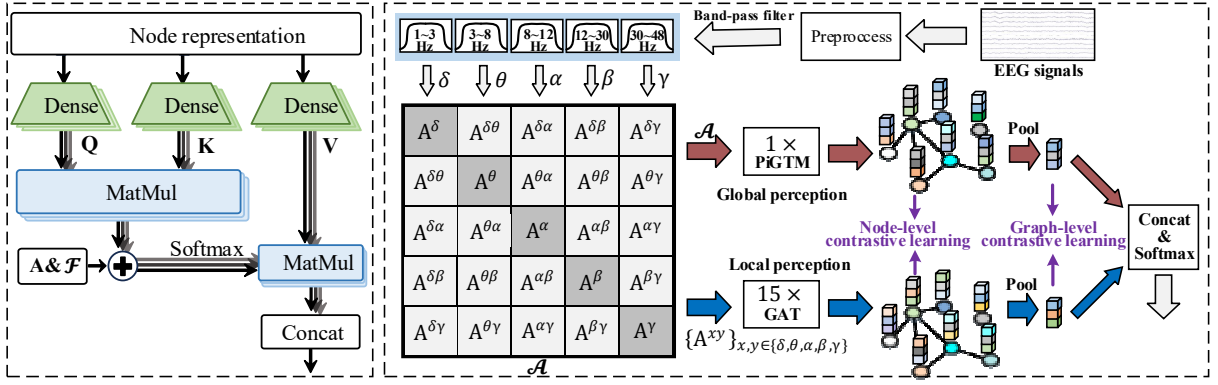

Fig. 2. Architecture of our proposed algorithm. (a) The structure of the self-attention mechanism in the prior information-based graph transformer module (PiGTM); (b) Dual-branch graph neural network (DB-GNN).

The attention matrix $\mathbf{E} \in \mathbb{R}^{n \times n}$ can be calculated as:

$$\mathbf{E} = \frac{\mathbf{Q}(\mathbf{K})^T}{\sqrt{d_K}} \tag{11}$$

Unlike the Graph Attention Network (GAT) [17], PiGTM not only uses the adjacency matrix for masking but also injects prior coupling information into the computation of self-attention, guiding the model's perception process and enhancing its generalization performance. The self-attention computation in GAT is as follows:

$$\mathbf{E} = \frac{\mathbf{Q}(\mathbf{K})^T}{\sqrt{d_K}} \odot \mathbf{A} \tag{12}$$

where **A** is the adjacency matrix of the brain network, serving as a sparse mask. $\odot$ represents element-wise multiplication. Next, we inject the coupling information into the attention matrix to enhance the model's perception capability. **E** is updated as:

$$\mathbf{E} = \frac{\mathbf{Q}(\mathbf{K})^T}{\sqrt{d_K}} \odot \mathbf{A} + \mathcal{F}, \mathcal{F} \leftarrow MLP(\mathbf{M}) \tag{13}$$

where $\mathbf{M} \leftarrow \left(\psi(m_{ij})\right) \in \mathbb{R}^{n \times n}$. $\psi: \mathbb{R} \rightarrow \mathbb{R}$ are learnable scalars indexed by $m_{ij}$. $m_{ij}$ is the degree of coupling between nodes $i$ and $j$ in the brain network from equation (2) and (3). Therefore, the self-attention output of PiGTM is

$$\text{Atten}(\mathbf{Q}, \mathbf{K}, \mathbf{V}) = MLP(\text{softmax}(\mathbf{E})\mathbf{V}) \tag{14}$$

Since PiGTM adopts the Transformer architecture, its final output **O** is:

$$\mathbf{O} = \text{LayerNorm}\left(\mathbf{Z} + \text{FFN}(\mathbf{Z})\right), \mathbf{Z}$$
$$\leftarrow \text{LayerNorm}\left(\mathbf{H} + \text{Atten}(\mathbf{Q}, \mathbf{K}, \mathbf{V})\right) \tag{15}$$

where FFN denotes the Feed-Forward Network, and **H** represents the original input features of the nodes.

*C. DB-GNN: Dual-Branch Graph Neural Network*

DB-GNN employs a dual-branch architecture: the global perception branch integrates information across all WFC (within-frequency coupling) and CFC (cross-frequency coupling) networks to extract global features from a holistic perspective, while the local perception branch focuses on capturing localized information within individual WFC or CFC networks, enhancing the model's ability to discern patterns from local structures. Furthermore, DB-GNN incorporates contrastive learning at both node and graph levels to regularize the training process of the dual-branch networks, thereby improving the model's generalization capability.

Regarding the global branch of perception, the input adjacency matrix is $\mathcal{A} \in \mathbb{R}^{(5 \times C) \times (5 \times C)}$ with 5 frequency bands and $C$ EEG sampling points. The mask matrix $\mathcal{A}$ and the prior coupling information $\mathcal{F}$ are used to guide the representation learning process of PiGTM (i.e., Eq. (13)). We then generate the global perception vector $\mathbf{H}^{global} = \left(\mathbf{h}_i^{global}\right)_{i=1,2,\dots,C} \in \mathbb{R}^{C \times d'}$ ($d'$ is the embedding dimension) by averaging these representations from the same node across frequency bands. For the local perception branch, we input 15 adjacency matrices $\mathbf{A}^{\delta}, \mathbf{A}^{\delta\theta}, \mathbf{A}^{\delta\alpha}, \mathbf{A}^{\delta\beta}, \mathbf{A}^{\delta\gamma}, \mathbf{A}^{\theta}, \mathbf{A}^{\theta\alpha}, \mathbf{A}^{\theta\beta}, \mathbf{A}^{\theta\gamma}, \mathbf{A}^{\alpha}, \mathbf{A}^{\alpha\beta}, \mathbf{A}^{\alpha\gamma}, \mathbf{A}^{\beta}, \mathbf{A}^{\beta\gamma}, \mathbf{A}^{\gamma}$ into GAT separately for representation learning, and run the average pooling of the 15 outputs from GAT to obtain the local perception vector of the network $\mathbf{H}^{local} = \left(\mathbf{h}_i^{local}\right) \in$ $\mathbb{R}^{C \times d'}$. To mitigate the risk of overfitting, DB-GNN adopts a graph contrastive learning approach to enforce dual-branch learning constraints. As shown in Fig. 3, since the dual-branch learning process extracts features from the same brain state, the feature vectors learned for the same node by the two branches are treated as positive samples. To enhance the differentiation between features of different nodes and improve the model's perception ability, feature vectors from other nodes are selected as negative samples. Given that this study focuses on the brain network classification task, we further impose feature constraints at the graph-level. As depicted in Fig. 3, DB-GNN first uses mean pooling to obtain graph-level features. Then, the global and local features of the same brain network learned by the dual-branch structure are treated as positive samples. Meanwhile, features from other brain networks in the same batch are selected as negative samples, thereby enhancing the graph-level feature perception ability of DB-GNN. Based on the InfoNCE loss [23], this study constructs the following contrastive learning loss functions for both node-level and graph-level features:

$$\mathcal{L}_{reg} = \mathcal{L}_{node} + \mathcal{L}_{graph} \quad (16)$$

$$\mathcal{L}_{node} = -\sum_{i=1}^{C}\left[\log\left(\frac{\exp\left(\frac{f(\mathbf{h}_i^{global},\mathbf{h}_i^{local})}{\tau}\right)}{\exp\left(\frac{f(\mathbf{h}_i^{global},\mathbf{h}_i^{local})}{\tau}\right)+\sum_{j=1}^{N}\frac{f(\mathbf{h}_i^{global},\mathbf{h}_j^{global})}{\tau}}\right)+\log\left(\frac{\exp\left(\frac{f(\mathbf{h}_i^{local},\mathbf{h}_i^{global})}{\tau}\right)}{\exp\left(\frac{f(\mathbf{h}_i^{local},\mathbf{h}_i^{global})}{\tau}\right)+\sum_{j=1}^{N}\frac{f(\mathbf{h}_i^{local},\mathbf{h}_j^{local})}{\tau}}\right)\right] \quad (17)$$

$$\mathcal{L}_{graph} = -\sum_{i=1}^{batchsize}\left[\log\left(\frac{\exp\left(\frac{f(\mathcal{H}_i^{global},\mathcal{H}_i^{local})}{\tau}\right)}{\exp\left(\frac{f(\mathcal{H}_i^{global},\mathcal{H}_i^{local})}{\tau}\right)+\sum_{j=1}^{N}\exp\left(\frac{f(\mathcal{H}_i^{global},\mathcal{H}_j^{global})}{\tau}\right)}\right)+\log\left(\frac{\exp\left(\frac{f(\mathcal{H}_i^{local},\mathcal{H}_i^{global})}{\tau}\right)}{\exp\left(\frac{f(\mathcal{H}_i^{local},\mathcal{H}_i^{global})}{\tau}\right)+\sum_{j=1}^{N}\exp\left(\frac{f(\mathcal{H}_i^{local},\mathcal{H}_j^{local})}{\tau}\right)}\right)\right] \quad (18)$$

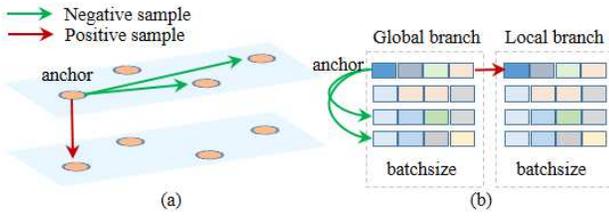

Fig. 3. The contrastive learning in DB-GNN. (a) Node-level contrastive learning; (b) Graph-level contrastive learning.

$\tau$ is the given temperature and $\mathcal{H}_i^{global}$ represents the global representation vector of the graph. Consequently, the loss function of the DB-GNN is:

$$\mathcal{L} = \mathcal{L}_{cls} + \epsilon \mathcal{L}_{reg} \quad (19)$$

with $\mathcal{L}_{graph}$ describing the classification loss of the brain network. DB-GNN combines the global and local features of the brain network and generates the classification result through a two-layer MLP. $\mathcal{L}_{cls}$ is computed by the categorical cross-entropy loss. $\epsilon$ is an artificially specified smaller parameter.

**Comments:** DB-GNN's dual-branch architecture employs PiGTM to globally capture the holistic topology of both CFC and WFC brain networks, while GAT extracts localized features from individual coupling networks, enabling multi-scale exploration of within-band neural couplings and cross-band oscillations in emotional processing. The PiGTM module (Eq.13) constrains the Transformer's representation learning process by embedding PLV/MI coupling strengths as domain priors, mitigating overfitting. Multi-level contrastive learning (Eq.16-18) explicitly optimizes collaborative representation learning in the dual-branch network through node- and graph-level constraints, guiding the model to learn complementary and discriminative coupling patterns. The experimental results validate DB-GNN's effectiveness in modeling the synergistic characteristics of within- and cross-frequency couplings in emotional brain networks.

## IV. EXPERIMENTS

### A. Experimental Details

This study utilizes the SEED dataset [27] containing 15 subjects to evaluate the proposed DB-GNN model. The SEED dataset classifies the emotional states into three categories – negative, positive, and neutral. Moreover, we use a three-second time window without overlapping to extract the PLV and MI features of the five frequency bands ($\delta, \theta, \alpha, \beta, \gamma$) based on multi-channel EEG signals. In terms of performance evaluation, we conducted experiments in a subject-dependent manner. 80% of the data is used as the training dataset, and the remaining 20% is used as the testing dataset.

In all experiments, we compare DB-GNN with several state-of-the-art (SOTA) baselines for graph classification: GCN [28], GAT [17], SuperGAT [29], AntiSymmetric [30] and pmlp [31]. This experiment adopts a three-layer graph learning structure and a global mean pooling method for baselines. The setting of hyperparameters mainly refers to [30], [31], and the grid search is utilized for parameter optimization. The Adam is adopted as the optimizer. The learning rate and (β1, β2) are set to 5e−4 and (0.9, 0.999). All models are trained for 300 steps on a NVIDIA A800 Tensor Core GPU.

### B. Verification Experiments

Table I and Table II present the testing accuracy and F1-scores of different algorithms across 15 subjects in the SEED dataset, respectively. The proposed DB-GNN model can achieve a mean testing accuracy of 97.88 ± 0.87 and an F1-score of 97.87 ± 0.87 to classify the positive, neutral, and negative emotions in subject-dependent experiments, which is significantly better and more stable than other baselines on the SEED dataset. This result demonstrates that the coordinated feature representation learning in both WFC and CFC brain networks and the multi-level graph contrastive learning can significantly improve the model performance on emotion recognition. It is noteworthy that the AntiSymmetric algorithm exhibited extremely low performance on subjects 7, 13, and 14. In contrast, the DB-GNN consistently maintained a testing accuracy above 95%. The variation between subjects can be caused by a variety of factors, including subjects' education background, sociability and their true evoked emotional state when participating in experiments. Therefore, the consistency of better performance of our model across different subjects suggests that DB-GNN possesses strong generalization performance.

TABLE I. TESTING ACCURACY OF DB-GNN AND SOTA BASELINES ON THE *SEED* DATASET.

| Subjects | GCN | GAT | SuperGAT | AntiSymmetric | DirGNN | pmlp | DB-GNN |
|---|---|---|---|---|---|---|---|
| 1 | 94.01 | 89.66 | 94.79 | 90.80 | 87.16 | 93.44 | 96.86 |
| 2 | 91.08 | 92.58 | 93.94 | 56.85 | 89.30 | 87.23 | 96.79 |
| 3 | 91.94 | 93.30 | 95.15 | 82.24 | 90.23 | 91.80 | 97.43 |
| 4 | 92.94 | 94.29 | 95.93 | 88.16 | 92.65 | 89.09 | 97.65 |
| 5 | 93.44 | 87.66 | 95.51 | 85.52 | 76.68 | 90.01 | 97.15 |
| 6 | 96.29 | 95.79 | 97.72 | 93.01 | 86.80 | 93.94 | 98.43 |
| 7 | 94.58 | 93.44 | 96.15 | 34.31 | 91.73 | 92.30 | 98.86 |
| 8 | 97.22 | 97.15 | 98.15 | 96.79 | 94.29 | 95.58 | 98.29 |
| 9 | 93.72 | 93.37 | 97.43 | 95.22 | 91.44 | 93.15 | 98.50 |
| 10 | 93.15 | 87.02 | 92.87 | 92.15 | 88.80 | 89.80 | 97.65 |
| 11 | 95.93 | 94.58 | 96.50 | 94.08 | 94.29 | 95.01 | 98.00 |
| 12 | 95.01 | 88.80 | 96.93 | 88.94 | 78.60 | 92.08 | 98.22 |
| 13 | 94.65 | 94.37 | 96.79 | 34.17 | 87.59 | 93.01 | 98.15 |
| 14 | 92.87 | 90.51 | 95.22 | 35.81 | 87.45 | 90.30 | 96.50 |
| 15 | 98.79 | 99.64 | 99.43 | 99.64 | 99.64 | 99.07 | 99.79 |
| Mean±Std. | 94.37±1.97 | 92.81±3.41 | 96.17±1.69 | 77.85±24.38 | 89.11±5.79 | 92.39±2.94 | **97.88±0.87** |

TABLE II. F1-SCORE OF DB-GNN AND SOTA BASELINES ON THE *SEED* DATASET.

| Subjects | GCN | GAT | SuperGAT | AntiSymmetric | DirGNN | pmlp | DB-GNN |
|---|---|---|---|---|---|---|---|
| 1 | 94.01 | 89.62 | 94.80 | 90.79 | 86.92 | 93.38 | 96.83 |
| 2 | 91.09 | 92.50 | 93.93 | 52.64 | 89.30 | 87.20 | 96.79 |
| 3 | 91.94 | 93.22 | 95.14 | 81.92 | 90.04 | 91.80 | 97.43 |
| 4 | 92.92 | 94.26 | 95.94 | 88.03 | 92.73 | 89.05 | 97.61 |
| 5 | 93.42 | 87.49 | 95.45 | 85.21 | 76.05 | 89.97 | 97.13 |
| 6 | 96.27 | 95.72 | 97.69 | 92.96 | 86.68 | 93.86 | 98.42 |
| 7 | 94.46 | 93.42 | 96.10 | 17.03 | 91.49 | 92.27 | 98.84 |
| 8 | 97.11 | 97.11 | 98.11 | 96.79 | 94.13 | 95.51 | 98.29 |
| 9 | 93.69 | 93.32 | 97.41 | 95.19 | 91.39 | 93.10 | 98.50 |
| 10 | 93.11 | 86.98 | 92.76 | 91.96 | 88.90 | 89.64 | 97.61 |
| 11 | 95.91 | 94.54 | 96.51 | 94.08 | 94.20 | 94.99 | 98.00 |
| 12 | 95.01 | 88.81 | 96.93 | 88.80 | 78.40 | 92.08 | 98.21 |
| 13 | 94.57 | 94.34 | 96.74 | 16.98 | 87.32 | 92.93 | 98.14 |
| 14 | 92.84 | 90.25 | 95.14 | 17.58 | 87.29 | 90.29 | 96.51 |
| 15 | 98.78 | 99.64 | 99.41 | 99.64 | 99.64 | 99.06 | 99.79 |
| Mean±Std. | 94.34±1.96 | 92.75±3.43 | 96.14±1.70 | 73.97±31.32 | 88.97±5.91 | 92.34±2.94 | **97.87±0.87** |

In Fig. 4, we employed the Wilcoxon rank-sum test to assess the significance of performance differences between DB-GNN and other baselines (***: $p < 0.001$, **: $p < 0.01$, *: $p < 0.05$). DB-GNN significantly outperforms all comparative algorithms, with its performance distribution being not only concentrated but also higher than those of other algorithms, demonstrating the advantages and robustness of DB-GNN in the task.

Fig. 5 presents the confusion matrices of various algorithms. Our algorithm effectively classifies the three emotions. In contrast, AntiSymmetric tends to misclassify positive and neutral emotions as negative emotions. DirGNN often misclassifies positive emotions as neutral ones. Overall, DB-GNN achieves superior classification of each emotion category, with only a minimal number of misclassifications.

Table III presents the recent studies on the SEED dataset. The DB-GNN proposed in this study achieves the highest testing accuracy and F1-score. Furthermore, the standard deviation (std) of DB-GNN's performance metrics across different subjects is relatively low, suggesting that our proposed method exhibits a certain level of generalization capability.

### C. Ablation Experiments

To verify the effectiveness of the presented architecture, we conducted ablation experiments on the SEED dataset, and the specific ablation results are shown in Table IV and Fig. 6. First, to verify the role of the dual-branch structure, we removed the global perception branch and only kept the local perception branch in the model (Model 1). The dramatic performance difference between Model 1 and DB-GNN demonstrates the importance of global features in model's recognition ability. Furthermore, to validate the usefulness of the coupling information (i.e., Eq. (13)), we added the global branch in Model 2 but omitted the prior coupling in formation $\mathcal{F}$. Model 3 contained the prior coupling information but lacked the regularization of graph contrastive learning as in DB-GNN. Therefore, by comparing Model 2 and Model 3, we conclude that the inclusion of prior coupling information in graph feature extraction is advantageous. Furthermore, by contrasting Model 3 and DB-GNN, we prove that the regularization of the learning process is indeed effective at enhancing the model's performance. Fig. 6 presents the confusion matrix of the algorithm. With the introduction of a dual-branch structure, prior coupling information, and graph contrastive learning-based regularization, the model's recognition capability gradually improves, particularly with a significant rise in the identification accuracy of positive samples and a reduction in the misclassification of positive samples as neural samples.

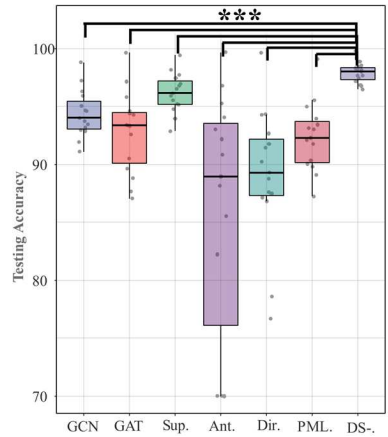

Fig. 4. Distribution of testing accuracy of SOTA baselines and DB-GNN in *SEED* dataset. Note: Sup., Ant., Dir., PML., and DS-. Represent SuperGAT, AntiSymmetric, DirGNN, pmlp and DB-GNN.

Fig. 5. Confusion matrices of SOTA baselines and DB-GNN on *SEED* dataset.

Fig. 6. Confusion matrices of comparison models on *SEED* dataset.

## V. CONCLUSION AND FUTURE DIRECTIONS

This study proposed a dual-branch graph neural network (DB-GNN) with multi-level contrastive learning to jointly identify WFC and CFC brain networks for emotion recognition. The dual-branch architecture simultaneously captures local and global features of brain networks, and the PiGTM enhances global feature perception while mitigating overfitting. Moreover, multi-level graph contrastive learning was employed to regularize the learning process, further improving the model's generalization capability. Experimental results on the SEED dataset demonstrated that DB-GNN achieved SOTA performance in emotion recognition tasks. Looking ahead, several directions warrant further exploration. The current model relies on predefined frequency bands and coupling metrics (e.g., PLV and MI). Future work could explore adaptive frequency band selection and more sophisticated coupling measures to better capture the dynamic nature of brain networks. Besides, the integration of additional modalities, such as functional magnetic resonance imaging (fMRI) or physiological signals, could further enhance emotion recognition performance with complementary information.

TABLE III. SOTA METHODS ON THE *SEED* DATASET.

| Studies | Year | Method | TeAcc±std | F1-score±std |
|---|---|---|---|---|
| Xu et al. [32] | 2019 | GIN | 88.64±2.26 | 88.63±2.25 |
| Zhong et al. [22] | 2022 | RGNN | 94.24±5.95 | --- |
| Jiang et al. [33] | 2023 | EmoGT | 95.02±5.99 | --- |
| Song et al. [34] | 2023 | V-IAG | 95.64±5.08 | --- |
| Li et al. [35] | 2023 | GMSS | 96.48±4.63 | --- |
| Pan et al. [36] | 2023 | MSFR-GCN | 96.63±4.64 | --- |
| Jeong et al. [37] | 2023 | HSCFLM | 97.20±10.1 | --- |
| Cai et al. [38] | 2024 | EEG-SWTNS | 94.83±7.16 | 94.27±2.79 |
| Gao et al. [39] | 2024 | CU-GCN | 95.70±5.32 | --- |
| Ju et al. [40] | 2024 | TDMNN | 97.20±1.57 | --- |
| Li et al. [41] | 2024 | BF-GCN | 97.44±3.89 | --- |
| Xue et al. [15] | 2024 | SAGN | 97.62±0.74 | 97.63±0.73 |
| **Ours** | **2025** | **DB-GNN** | **97.88±0.87** | **97.87±0.87** |

TABLE IV. ABLATION STUDIES ON THE *SEED* DATASET.

| Subjects | Model 1 | Model 2 | Model 3 | DB-GNN |
|---|---|---|---|---|
| 1 | 90.16 | 90.73 | 95.44 | 96.83 |
| 2 | 85.02 | 88.37 | 92.87 | 96.79 |
| 3 | 91.23 | 91.44 | 94.94 | 97.43 |
| 4 | 91.80 | 92.37 | 94.58 | 97.61 |
| 5 | 88.73 | 88.94 | 94.08 | 97.13 |
| 6 | 94.86 | 95.22 | 97.08 | 98.42 |
| 7 | 93.44 | 92.94 | 96.01 | 98.84 |
| 8 | 96.01 | 95.93 | 97.72 | 98.29 |
| 9 | 94.08 | 95.22 | 96.36 | 98.50 |
| 10 | 88.87 | 89.73 | 93.94 | 97.61 |
| 11 | 94.29 | 96.29 | 97.50 | 98.00 |
| 12 | 90.87 | 93.51 | 95.79 | 98.21 |
| 13 | 87.52 | 94.72 | 96.01 | 98.14 |
| 14 | 87.30 | 92.87 | 94.51 | 96.51 |
| 15 | 99.22 | 99.22 | 99.43 | 99.79 |
| **Mean±Std.** | 91.56±3.67 | 93.17±2.91 | 95.75±1.65 | **97.87±0.87** |

## ACKNOWLEDGMENT

This research was partly supported by research grants of Shenzhen-Hong Kong-Macau S&T Program (Category C) (SGDX20201103095002019), Shenzhen Basic Research Program (JCYJ20210324130209023), Mainland-Hong Kong Joint Funding Scheme (MHKJFS) (MHP/005/20), 'Chunhui plan' Project Foundation of the Education Department of China (HZKY20220133, 202200461), Project of Strategic Importance Fund (P0035421) and Project of RI-IWEAR fund(P0038684) from The Hong Kong Polytechnic University, and by High-end Foreign Experts Recruitment Plan of China (H20240949), and by the national natural science foundation of Jiangsu, China under Grant BK20241831, and by the natural science foundation of Jiangsu Universities, China under Grant 19JKD520003, and by the national defense basic research program of China under Grant JCKY2020206B037, and by National Key Laboratory of Ship Structural Safety (459900925) and by Postgraduate Research & Practice Innovation Program of Jiangsu Province under Grant KYCX21_3506, KYCX22_3825 and KYCX23_3890.